\documentclass[12pt,english,letter]{article}
\pdfoutput=1
\usepackage[T1]{fontenc}
\usepackage{graphicx,array}
\usepackage{cite}
\usepackage{color}
\usepackage{dsfont}
\usepackage{latexsym}
\usepackage{amsthm}
\usepackage{amsmath}
\usepackage{stackengine}
\usepackage[titletoc]{appendix}
\usepackage{enumitem}
\usepackage{amssymb}
\usepackage{color}
\usepackage[unicode=true,
 bookmarks=true,bookmarksnumbered=false,bookmarksopen=false,
 breaklinks=false,pdfborder={0 0 1},backref=false,colorlinks=true]
 {hyperref}
\hypersetup{pdftitle={Charged Dilatonic Spacetimes in String Theory},
 pdfauthor={Achilleas P. Porfyriadis, Grant N. Remmen},
 citecolor=black,linkcolor=black,urlcolor=black}
\usepackage{breakurl}
\usepackage[hang,flushmargin]{footmisc}

\newcolumntype{L}[1]{>{\raggedright\let\newline\\\arraybackslash\hspace{0pt}}m{#1}}
\newcolumntype{C}[1]{>{\centering\let\newline\\\arraybackslash\hspace{0pt}}m{#1}}

\def\simgt{\mathrel{\lower2.5pt\vbox{\lineskip=0pt\baselineskip=0pt
           \hbox{$>$}\hbox{$\sim$}}}}
\def\simlt{\mathrel{\lower2.5pt\vbox{\lineskip=0pt\baselineskip=0pt
           \hbox{$<$}\hbox{$\sim$}}}}

\newcommand{\be}{\begin{equation}}
\newcommand{\ee}{\end{equation}}
\newcommand{\bea}{\begin{eqnarray}}
\newcommand{\eea}{\end{eqnarray}}

\newcommand*\oline[1]{%
  \vbox{%
    \hrule height 0.5pt
    \kern0.68ex
    \hbox{%
      \kern-0.1em
      \ifmmode#1\else\ensuremath{#1}\fi
      \kern-0.1em
    }
  }
}

\definecolor{nicered}{rgb}{0.7,0.1,0.1}
\definecolor{nicegreen}{rgb}{0.1,0.5,0.1}
\hypersetup{
 colorlinks,citecolor=black,linkcolor=black,urlcolor=black}
\usepackage{xcolor}

\renewcommand{\Box}{\square}

\begin{document}

\interfootnotelinepenalty=10000
\baselineskip=18pt
\hfill CCTP-2023-1 

\hfill ITCP-IPP 2023/1

\vspace{2cm}
\thispagestyle{empty}
\begin{center}
{\LARGE \bf
Charged Dilatonic Spacetimes in String Theory
}\\
\bigskip\vspace{1cm}{
{\large Achilleas P. Porfyriadis${}^{a}$ and Grant N. Remmen${}^{b}$}
} \\[7mm]
{
\it 
${}^a$Crete Center for Theoretical Physics, Institute of Theoretical and Computational Physics, Department of Physics, University of Crete, 70013 Heraklion, Greece\\[1.5 mm]
${}^b$Kavli Institute for Theoretical Physics and Department of Physics, \\ University of California, Santa Barbara, CA 93106, United States
}
\let\thefootnote\relax\footnote{e-mail: 
\url{porfyriadis@physics.uoc.gr}, \url{remmen@kitp.ucsb.edu}}
 \end{center}

\bigskip
\centerline{\large\bf Abstract}
\begin{quote} \small
We construct and study general static, spherically symmetric, magnetically charged solutions in Einstein-Maxwell-dilaton gravity in four dimensions.
That is, taking Einstein gravity coupled to a ${\rm U}(1)$ gauge field and a massless dilaton---e.g., the action in the low-energy limit of string theory or Kaluza-Klein reduction---with arbitrary dilaton coupling, we build a three-parameter family of objects characterized by their mass, charge, and dilaton flux, generalizing the well known Garfinkle-Horowitz-Strominger black hole.
We analyze the near-extremal and near-horizon behavior in detail, finding new warped geometries.
In a particular limit, where the geometry reduces to the recently discovered customizable ${\rm AdS}_2 \times S^2$ of Einstein-Maxwell-dilaton gravity, we compute the static s-wave linearized solutions and characterize the anabasis relating the horizon perturbations to their nonlinear completions within our generalized family of spacetimes.
\end{quote}
	
\setcounter{footnote}{0}

\newpage
\tableofcontents
\newpage

\section{Introduction}
\label{sec:intro}

For a nonlinear system, the identification and study of highly symmetric nonperturbative solutions, if they can be found, offers a crucial window into its fundamental physical attributes, both classical and quantum.
In the case of Einstein-Maxwell-dilaton (EMD) theory, describing gravity coupled to a Maxwell field and a massless dilaton, the crucial objects of study are the black holes, found by investigating static, spherically symmetric, charged solutions.
In their seminal 1991 work~\cite{GHS},\footnote{See also work by Gibbons~\cite{Gibbons1982} and Gibbons and Maeda~\cite{GibbonsMaeda1988}.} Garfinkle, Horowitz, and Strominger (GHS) identified charged black hole solutions in string theory, solving the equations of motion for the EMD theory, which in string frame is described by the Lagrangian,
\be
{\cal L} = e^{-2\lambda\phi} \left[R + 2(3\lambda^2 - 1)\nabla_\mu \phi \nabla^\mu \phi - \frac{1}{2}F_{\mu\nu}F^{\mu\nu}  \right].\label{eq:EMD} 
\ee
Here, we have let the dilaton coupling $\lambda$ be arbitrary.
The choice $\lambda=1$ corresponds to the low-energy effective action of the heterotic string, while one obtains $\lambda=\sqrt{3}$ from Kaluza-Klein compactification of five-dimensional gravity.
In terms of the Einstein-frame metric, where we send $g_{\mu\nu} \rightarrow e^{-2\lambda\phi} g_{\mu\nu}$, the Lagrangian is given simply by $R - 2(\nabla\phi)^2 - \frac{1}{2}e^{-2\lambda\phi}F^2$, but we will find it calculationally more convenient to work in string frame throughout unless otherwise noted. Neither string nor Einstein frame is a priori more or less physical than the other: while the behavior of geodesics and curvatures are different between the two frames, which one a given observer experiences is determined by whether they are composed of matter that is minimally coupled to the string- or Einstein-frame metric. Indeed, in string theory, the former generically holds.

In Ref.~\cite{GHS}, GHS found a family of magnetic solutions for arbitrary $\lambda$.
Like their Reissner-Nordstr\"om predecessors, GHS black holes are characterized by two parameters: the mass $M$ and (magnetic) charge $P$.
However, while in Einstein-Maxwell theory, Birkhoff's theorem guarantees uniqueness of the Reissner-Nordstr\"om solution as the only spherically symmetric, asymptotically flat electrovacuum, this result does not apply in the presence of a massless dilaton.
That is, we can ask whether the equations of motion  for the theory in Eq.~\eqref{eq:EMD} allow for dilatonic hair, describing a {\it three}-parameter family of spacetimes characterized by $M$, $P$, and a new charge $D$ measuring the integrated dilaton flux at infinity. In such an extended family of objects, the GHS solution would occupy a particular two-dimensional slice through parameter space.

In Refs.~\cite{Rakhmanov:1993yd,Turimov:2021uej}, the $\lambda\,{=}\,1$ GHS solutions were extended to a three-parameter family, but without providing explicit general solutions for arbitrary dilaton coupling $\lambda$.\footnote{See also Ref.~\cite{Yazadjiev:2000sa} for a procedure relating solutions in Einstein gravity to ones in the EMD theory.}
In other works, various exotic EMD systems and their solutions have been studied, including adding a cosmological constant, dilaton potentials, higher dimensions, rotation, Yang-Mills fields, higher-derivative gravity, and more, which we will not attempt to catalogue here.
General static, spherically symmetric, electric  EMD solutions for arbitrary dilaton coupling were constructed and categorized in Refs.~\cite{Gurses:1995fw,Nozawa:2020wet},\footnote{While Ref.~\cite{Gurses:1995fw} found that only the GHS black hole was regular in Einstein frame, we will find a somewhat different situation in string frame.} but a consideration of the family of spacetimes in string frame, and in particular a thorough study of their extremal limits, is warranted.

This paper is organized as follows.
In Sec.~\ref{sec:general} we present the general construction for static, spherically symmetric solutions in EMD gravity, finding the three-parameter set of asymptotically flat spacetimes in Eq.~\eqref{eq:solg} and exploring the near-horizon scaling and smoothness of the metric.
We compute the asymptotic charges $(M,P,D)$ and characterize the extremality bound in Sec.~\ref{sec:charges}.
In Sec.~\ref{sec:extremal}, we analyze the near-horizon, near-extremal (NHNE) limit of the metric; while we demonstrated an ${\rm AdS}_2 \times S^2$ spacetime emerging from GHS in Ref.~\cite{Porfyriadis:2021zfb}, for our more general EMD solutions here, a richer NHNE geometry arises in Eq.~\eqref{eq:nearhorizon}.

In contrast to the nonlinear solutions of the previous sections, in Sec.~\ref{sec:pert} we instead analyze the linearized EMD equations in the ${\rm AdS}_2 \times S^2$ background.
While we found the general set of propagating modes in Ref.~\cite{Porfyriadis:2021zfb}, here we instead consider the static s-wave sector, which must be analyzed separately.
Modulo diffeomorphisms, along with a mode corresponding to an overall rescaling of the geometry, we find three physically distinct linearized solutions.
In Sec.~\ref{sec:anabasis}, we connect our linearized analysis to our nonnlinear solutions of Secs.~\ref{sec:general} and \ref{sec:extremal}, solving the anabasis problem\footnote{Following Ref.~\cite{Hadar:2020kry}, which considered the analogous problem for Reissner-Nordstr\"om black holes, we use the nomenclature {\it anabasis} in its historical sense of a march up-country from a boundary to the interior, in this case, bootstrapping a bulk geometry from its behavior near the horizon.} for our charged dilatonic objects by identifying the three linearized solutions as first-order deviations away from the near-horizon limit.
We conclude and discuss open questions in Sec.~\ref{sec:conclusions}.

\section{General construction}\label{sec:general}

We wish to find the general static, spherically symmetric, magnetic solutions in EMD theory~\eqref{eq:EMD}.
The equations of motion comprise the Einstein equation,
\be 
\begin{aligned}
&&R_{\mu\nu} - \frac{1}{2}R g_{\mu\nu} - F_{\mu\rho} F_\nu^{\;\;\rho}  + \frac{1}{4}g_{\mu\nu} F_{\rho\sigma}F^{\rho\sigma} - 2(1-\lambda^2)\nabla_\mu \phi\nabla_\nu\phi  \\ && + (1+\lambda^2)g_{\mu\nu}\nabla_\rho \phi \nabla^\rho\phi + 2\lambda(\nabla_\mu \nabla_\nu \phi - g_{\mu\nu}\Box \phi) &= 0,
\end{aligned}\label{eq:ein}
\ee
the Maxwell equation,
\be 
\nabla_\mu F^{\mu\nu}  - 2\lambda F^{\mu\nu}\nabla_\mu \phi = 0,\label{eq:max}
\ee
and the Klein-Gordon equation,
\be 
2(3\lambda^2 - 1)\Box \phi  + \lambda \left[R - 2(3\lambda^2 - 1)\nabla_\mu\phi\nabla^\mu \phi - \frac{1}{2}F_{\mu\nu}F^{\mu\nu} \right] = 0.\label{eq:KG}
\ee
It will be useful notation to write the left-hand sides of the Einstein and Klein-Gordon equations as $E_\mu^{\;\;\nu}$ and $S$, respectively.
A benefit of the conformal frame defining our Lagrangian in Eq.~\eqref{eq:EMD} is that, in this parameterization, the equations of motion are purely derivatively coupled in the dilaton $\phi$, which does not hold for other choices (e.g., Einstein frame).
Since the equations of motion are invariant under swapping $\lambda\rightarrow -\lambda$ and $\phi\rightarrow -\phi$, we will take $\lambda>0$ without loss of generality.

We start with a general static, spherically symmetric ansatz for the metric in four spacetime dimensions, 
\be 
{\rm d}s^2 = -f(r) {\rm d}t^2 + \frac{{\rm d}r^2}{g(r)} + h(r) {\rm d}\Omega^2,\label{eq:ansatz}
\ee
along with $\phi = \phi(r)$ and a gauge field strength corresponding to a magnetic charge,
\be 
F = P\sin\theta\,{\rm d}\theta\wedge {\rm d}\varphi.\label{eq:F}
\ee
With this ansatz for $F$, the Maxwell equation \eqref{eq:max} is automatically satisfied.

We seek the full set of asymptotically flat solutions, where as $r\rightarrow\infty$, $f$, $g$, and $\phi$ approach constants and $h(r)\rightarrow r^2$.
Let us fix a gauge in which
\be
e^{-2\lambda \phi(r)} \frac{r^2}{h(r)}\sqrt{\frac{f(r)}{g(r)}} = {\rm constant},
\ee
where we define the constant to be $e^{-2\lambda\phi_0}$.
We can enforce this gauge via a diffeomorphism on $r$, and we see that it is consistent with asymptotic flatness.
Let us further define $j(r) = g(r)h(r)^2$ and eliminate $g$.

A priori, there will be four remaining independent equations of motion, $E_t^{\;\;t}$, $E_r^{\;\;r}$, $E_\theta^{\;\;\theta}$, and $S$.
We find that the particular combination
\be
E_t^{\;\;t} - \frac{1+\lambda^2}{1-\lambda^2}(E_r^{\;\;r} + 2 E_{\theta}^{\;\;\theta}) - \frac{\lambda}{1-\lambda^2} S = 0
\ee
depends on $j$ alone,
\be 
r^2 j''(r) - 6 r j'(r)+12 j(r) = (1+\lambda^2)P^2 r^2,
\ee
to which we have the general solution
\be
j(r) = \frac{1+\lambda^2}{2}P^2 r^2 + c_3 r^3 + c_4 r^4 .
\ee
We choose to write 
\be
P = \sqrt{\frac{2r_+ r_-}{1+\lambda^2}} \label{eq:P}
\ee
and $c_3 = -(r_+ + r_-)$ for some parameters $r_\pm$. This can be done without loss of generality since we are simply exchanging two parameters $P$ and $c_3$ for $r_\pm$, and $P$ enters the Einstein and Klein-Gordon equations only as $P^2$.~\footnote{While the choice in Eq.~\eqref{eq:P} is inspired by the GHS solution, and $r_\pm$ will eventually become horizons, this is not an input assumption.}
We then find that the combination
\be 
2\lambda E_t^{\;\;t} + S = 0
\ee
depends only on $f$ (and the known function $j(r)$):
\be
r^2 f(r)\frac{{\rm d}}{{\rm d} r}\left(\frac{j(r)}{r^2}f'(r)\right)- j(r) f'(r)^2 -\frac{2(1-\lambda^2)}{1+\lambda^2} r_+ r_- f(r)^2=0.
\ee
We have the general solution
\be 
f(r) = \frac{c_2 j(r)}{r^4}\left[\frac{2c_4 - r_+ - r_- + \sqrt{(r_1 + r_2)^2 - 4c_4 r_+ r_-}}{2c_4 - r_+ - r_- - \sqrt{(r_1 + r_2)^2 - 4c_4 r_+ r_-}}\right]^{c_1}
\ee
for some new constants $c_{1,2}$.
The equation $E_\theta^{\;\;\theta} = 0$ then implies
\be 
\begin{aligned}
&\left(\frac{j(r)}{r^2}\right)^2[8h(r)(2h'(r)+r h''(r))-12 r h'(r)^2]  \\ &\; + \frac{r}{\lambda^2(1+\lambda^2)}[r_+^2 + r_-^2 + 2(1-2c_4)r_+ r_-][c_1^2(1+\lambda^2)^2 - 4\lambda^4]h(r)^2 =0 .\label{eq:heq}
\end{aligned}
\ee
Expanding this equation of motion around $r=\infty$ and imposing $h\rightarrow r^2$ there for asymptotic flatness, we find that $c_4 = 1$.
We fix the constant $c_1$ using the remaining equations of motion and $c_2$ by a rescaling of $t$.
The general solution for $h$ from Eq.~\eqref{eq:heq} gives two free constants, one of which can be fixed using $E_r^{\;\;r}$. 
When the dust settles, we arrive at the general solution to the full set of EMD equations of motion in terms of three parameters: $r_+$, $r_-$, and a new constant $q$.

Explicitly, we find that the general solution for the geometry is
\be
\boxed{
\begin{aligned}
{\rm d}s^2 =&\, -\left[\frac{(r-r_+)(r-r_-)}{r^2}\right]^{\frac{1-\lambda^2}{1+\lambda^2}} \left(\frac{r-r_+}{r-r_-}\right)^{\pm \frac{2\lambda \sqrt{1-q^2+\lambda^2}}{q(1+\lambda^2)}}{\rm d}t^2  \\ &\, + \left(\frac{r_+ - r_-}{qr}\right)^4 \left[\frac{r^2}{(r-r_+)(r-r_-)}\right]^3\left[\left(\frac{r-r_+}{r-r_-}\right)^{\frac{1}{2q}} - \left(\frac{r-r_+}{r-r_-}\right)^{-\frac{1}{2q}}\right]^{-4} {\rm d}r^2 \\&\, + \frac{r^2 (r_+ - r_-)^2}{q^2 (r-r_+)(r-r_-)}\left[\left(\frac{r-r_+}{r-r_-}\right)^{\frac{1}{2q}} - \left(\frac{r-r_+}{r-r_-}\right)^{-\frac{1}{2q}}\right]^{-2}{\rm d}\Omega^2,
\end{aligned}}\label{eq:solg}
\ee
with corresponding dilaton profile
\be 
\phi = -\frac{\lambda}{2(1+\lambda^2)}\log\left[\frac{(r-r_+)(r-r_-)}{r^2}\right] \pm \frac{\sqrt{1-q^2+\lambda^2}}{2q(1+\lambda^2)}\log\left(\frac{r-r_+}{r-r_-}\right) + \phi_0.\label{eq:solphi}
\ee
For general $q$, for the geometry to remain Lorentzian we require 
\be
q^2 \leq 1+\lambda^2.\label{eq:reality}
\ee
We see above that swapping the sign in the $\pm$ in Eqs.~\eqref{eq:solg} and \eqref{eq:solphi} is equivalent to swapping $q\rightarrow -q$, so without loss of generality we will set $\pm \rightarrow +$ hereafter and further take $r>0$.
By the symmetry in $r_\pm$, we take $r_+>r_-$ without loss of generality.
To avoid the curvature singularity at $r=0$ (for example, $R \propto 1/r^4$), we impose $r_+>0$ and focus on the region $r>r_+$.

For $q=1$, the solution in Eqs.~\eqref{eq:solg} and \eqref{eq:solphi} reduces to the GHS black hole,
\be 
\begin{aligned}
{\rm d}s^2 &= -\left(1-\frac{r_+}{r}\right)\left(1-\frac{r_-}{r}\right)^{\frac{1-3\lambda^2}{1+\lambda^2}} {\rm d} t^2 + \left(1-\frac{r_+}{r}\right)^{-1}\left(1-\frac{r_-}{r}\right)^{-1}{\rm d}r^2 + r^2 {\rm d}\Omega^2\\
\phi &= \phi_0 - \frac{\lambda}{1+\lambda^2}\log\left(1-\frac{r_-}{r}\right).
\end{aligned}
\ee
Remarkably, {\it independent} of $q$, in the extremal limit where the two horizons are degenerate, $r_\pm \rightarrow r_0$, the general solution reduces to extremal GHS:
\be 
\begin{aligned}
{\rm d}s^2 &= -\left(1-\frac{r_0}{r}\right)^{\frac{2(1-\lambda^2)}{1+\lambda^2}} {\rm d} t^2 + \left(1-\frac{r_0}{r}\right)^{-2}{\rm d}r^2 + r^2 {\rm d}\Omega^2\\
\phi &= \phi_0 - \frac{\lambda}{1+\lambda^2}\log\left(1-\frac{r_0}{r}\right).\label{eq:extremal}
\end{aligned}
\ee
That is, the extremal GHS solution functions as a sort of attractor, in the $r_-\rightarrow r_+$ limit, for the general $q$-deformed family of non-GHS objects in Eq.~\eqref{eq:solg}.\footnote{We use the nomenclature ``$q$-deformed'' here in reference to the field of $q$-analogues in mathematics, in which familiar functions can be deformed via a parameter $q$ under which the original function is recovered when $q\rightarrow 1$ (e.g., the $q$-deformed logarithm $\log_q x = (x^{1-q}-1)/(1-q)$), in analogy with how we recover the GHS black hole in the $q\rightarrow 1$ limit of Eq.~\eqref{eq:solg}. Indeed, the function $x^{1/2q} - x^{-1/2q}$ for $x=(r-r_+)/(r-r_-)$, which appears in Eq.~\eqref{eq:solg}, can be written as $q^{-1}x^{\frac{1}{2q}}\log_{1+q^{-1}}(x)$.}
We will discuss subtleties of the extremal limit in much more detail in Sec.~\ref{sec:extremal}.

\begin{figure}[t]
\begin{center}
\includegraphics[width=8cm]{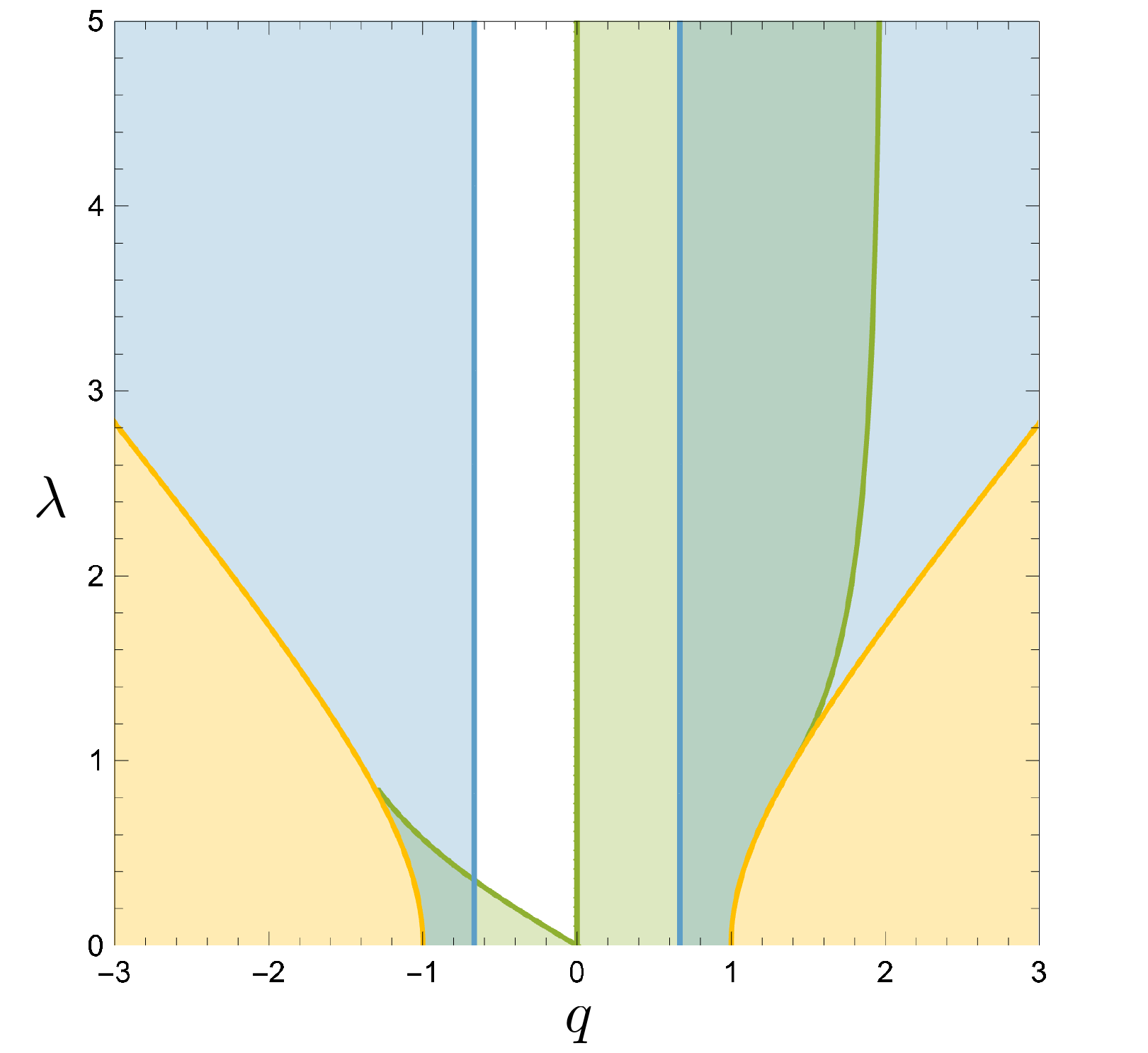}
\end{center}
\vspace{-0.5cm}
\caption{Conditions on $\lambda$ and $q$ necessary for the existence of a horizon in the string-frame metric given in Eq.~\eqref{eq:solg}. We have that $g_{tt}\propto f(r)$ and $g^{rr}\propto g(r)$ vanish in the $r\rightarrow r_+$ limit in the regions shaded green and blue, respectively. The yellow region is forbidden by Eq.~\eqref{eq:reality}, which enforces $q^2 \leq 1+\lambda^2$ for reality of the metric.}
\label{fig:horizon}
\end{figure}

The general solution is asymptotically flat. That is, taking the $r\rightarrow \infty$ limit, Eq.~\eqref{eq:solg} reduces to the Minkowski metric and $\phi\rightarrow \phi_0$.
A necessary condition for a horizon to exist at $r= r_+$ is if $f$ and $g$ both scale with positive powers of $(r-r_+)$ in the limit.
When $r_+ \neq r_-$, we find that $g(r)\propto (r-r_+)^{3-\frac{2}{|q|}}$, while $f(r)\propto (r-r_+)^{\frac{1-\lambda^2}{1+\lambda^2} + \frac{2\lambda\sqrt{1-q^2+\lambda^2}}{q(1+\lambda^2)}}$, giving rise to an interesting set of conditions; see Fig.~\ref{fig:horizon}.
In the extremal case where $r_- \rightarrow r_+$, the condition for the existence of the horizon simply becomes $\lambda < 1$.
We can explore properties of the $r=r_+$ surface by computing the Riemann tensor and transforming to the orthonormal frame of a static observer via $R^a_{\;\;bcd} = e^a_\mu e_b^\nu e_c^\rho e_d^\sigma R^\mu_{\;\;\nu\rho\sigma}$, where the vierbein is defined via $g_{\mu\nu} = e^a_\mu e^b_\nu \eta_{ab}$, for $\eta_{ab}={\rm diag}(-1,1,1,1)$.
In the GHS case ($q=1$), the curvature components of the string-frame metric are finite in the static orthonormal frame and the horizon is regular~\cite{GHS}. When $q\neq 1$, the nonzero components go like $(r-r_+)^{1-\frac{2}{q}}$, implying that the $r\rightarrow r_+$ limit is singular when $q<2$. However, when $q>2$, the orthonormal-frame curvature vanishes in the limit, and from the $f$ and $g$ scaling depicted in Fig.~\ref{fig:horizon}, the $r= r_+$ surface is infinitely far away, so that we enter another asymptotically flat region as $r\rightarrow r_+$.
In string frame, the area of the object is given by computing
\be
g_{\theta\theta} \stackrel{r\rightarrow r_+}{\longrightarrow}\frac{r_+^2}{q^2} \left(\frac{r-r_+}{r_+ - r_-}\right)^{-1+\frac{1}{q}}.
\ee
Thus, in string frame, the area vanishes when $q<1$ and diverges when $q>1$, remaining finite only in the GHS case.
In contrast, for the {\it Einstein frame} metric $g_{\mu\nu}^E = e^{-2\lambda\phi}g_{\mu\nu}$, setting the constant $\phi_0=0$, one finds
\be
g_{\theta\theta}^E \stackrel{r\rightarrow r_+}{\longrightarrow}\frac{(r_+-r_-)^2}{q^2}\left(\frac{r_+}{r_+-r_-}\right)^{\frac{2}{1+\lambda^2}} \left(\frac{r-r_+}{r_+ - r_-}\right)^{\frac{1}{q} - \frac{1}{1+\lambda^2} + \frac{\lambda\sqrt{1-q^2+\lambda^2}}{q(1+\lambda^2)}},
\ee
which goes to infinity if $q<0$ and zero if $q>0$, except for the $q=1$ GHS case where it is finite in the subextremal case.
That is, in Einstein frame, our $q$-deformed solutions are all pointlike objects when $q>0$, even when $r_+ \neq r_-$.

From this perspective, our construction is usefully thought of as characterizing the geometry outside a general charged object in string theory, without assuming the tuning between the dilatonic and electromagnetic flux present in the GHS black hole. 
We will consider charges in more detail in the next section.

\section{Charges}\label{sec:charges}
\begin{figure}[t]
\begin{center}
\includegraphics[width=8cm]{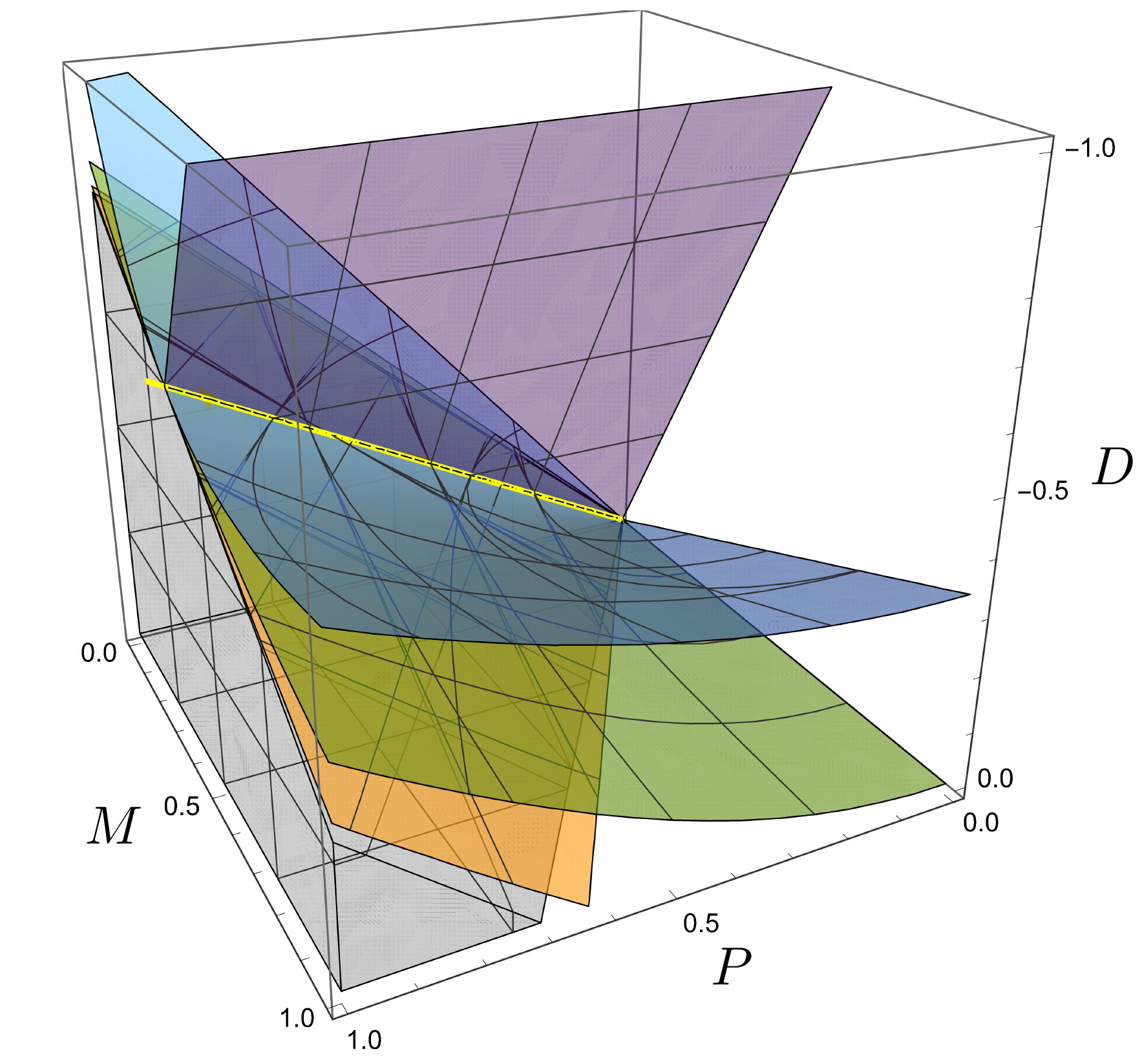}
\end{center}
\vspace{-0.5cm}
\caption{Characterization of the $q$-deformed charged, dilatonic geometry of Eq.~\eqref{eq:solg} in terms of mass $M$, magnetic charge $P$, and dilatonic charge $D$ defined in Eq.~\eqref{eq:charges}. In this plot, $\lambda=\sqrt{3}$, as one obtains under Kaluza-Klein reduction of five-dimensional gravity on a circle, and we have plotted the surfaces occupied by spacetimes for $q=1/2$ (orange), $q=1$ (green, GHS), $q=3/2$ (blue), and the maximal value of $q=\sqrt{1+\lambda^2} = 2$ (purple). The region forbidden by the extremality bound in Eq.~\eqref{eq:extremality} is shown in gray, and the $q$-independent universal extremal solution of Eq.~\eqref{eq:extremal} is given by the yellow line.}
\label{fig:MPD}
\end{figure}

We can better understand and characterize the family of spacetimes that we found in Eqs.~\eqref{eq:solg} and \eqref{eq:solphi} by defining asymptotic charges.
As all of these solutions are static and asymptotically flat, ADM charges can be computed equivalently using the Komar formulas.
For comparison with the literature, in this section we will implicitly use the Einstein-frame metric $e^{-2\lambda\phi}g_{\mu\nu}$ defined previously to contract all metrics and define all covariant derivatives (setting the constant $\phi_0$ to zero as before).
We write the Komar mass, magnetic charge, and integrated dilaton flux as follows:
\be
\begin{aligned}
M &= \frac{1}{4\pi}\int_{i^0} n_\mu \sigma_\nu \nabla^\mu K^\nu \,{\rm d}A\\
P &= \frac{1}{4\pi}\int_{i^0} n_\mu \sigma_\nu \widetilde F^{\mu\nu}\,{\rm d}A \\
D &= \frac{1}{4\pi}\int_{i^0} \sigma_\mu \nabla^\mu \phi\,{\rm d}A,
\end{aligned} 
\ee
where all integrals are taken over a sphere near spatial infinity $i^0$, the area element is ${\rm d}A = e^{-2\lambda\phi} g_{\theta\theta} \sin\theta\, {\rm d}\theta\,{\rm d}\phi$, and we have defined a unit timelike normal $n^\mu$, unit outward-pointing spacelike normal $\sigma^\mu$, the timelike Killing vector $K^\mu$, and the dual field strength tensor $\widetilde F^{\mu\nu} = \epsilon^{\mu\nu\rho\sigma}F_{\rho\sigma}/2$.
We find the charges
\be 
\begin{aligned}
M &= \frac{r_+ + r_-}{2(1+\lambda^2)} + \frac{\lambda(r_+ - r_-)}{2(1+\lambda^2)}\frac{\sqrt{1-q^2+\lambda^2}}{q}\\
P &= \sqrt{\frac{2 r_+ r_-}{1+\lambda^2}}\\
D &= -\frac{\lambda(r_+ + r_-)}{2(1+\lambda^2)} + \frac{r_+ - r_-}{2(1+\lambda^2)}\frac{\sqrt{1-q^2+\lambda^2}}{q}.
\end{aligned}\label{eq:charges}
\ee 
We see that the ADM definition of $P$ is consistent with Eq.~\eqref{eq:P}, as required.
Notably, we can understand the $q$ parameter describing deviation of the solution from the GHS black hole in terms of the charges alone,
\be
q^2 = \frac{2(M - \lambda D)^2-(1+\lambda^2)P^2}{2(M^2 + D^2)-P^2}. \label{eq:qdef}
\ee
In the three-dimensional $(M,P,D)$ space of charges, the GHS solution corresponds to the two-dimensional $q=1$ surface; see Fig.~\ref{fig:MPD}.
We note that the GHS case enjoys the distinction of having the unique $q$ value for which taking the neutral limit $P\rightarrow 0$ at fixed $r_+$ (i.e., $r_-\rightarrow 0$) implies vanishing dilaton flux $D\rightarrow 0$.
In terms of the charges, we can write the asymptotic form of the string-frame solution in Eqs.~\eqref{eq:solg} and \eqref{eq:solphi} as
\be
\begin{aligned}
{\rm d}s^2 &= -\left[1 - \frac{2(M+\lambda D)}{r}+\cdots\right]{\rm d}t^2 +\left[1 + \frac{2(M-\lambda D)}{r}+\cdots\right]{\rm d}r^2 + r^2 \left[1 +\cdots \right]{\rm d}\Omega^2  \\
\phi &= \phi_0 - \frac{D}{r} + \cdots
\end{aligned}
\ee
where $+\cdots$ indicates terms that fall off at least as fast at $O(1/r^2)$.

Furthermore, we find the extremality condition---which holds independent of $q$---by observing that
\be 
2(M-\lambda D)^2 - (1+\lambda^2) P^2 = \frac{(r_+ - r_-)^2}{2} \geq 0.\label{eq:extremality}
\ee
By Eq.~\eqref{eq:qdef}, one also finds that $2(M^2+D^2)-P^2 = (r_+-r_-)^2/2q^2 \geq 0$, but this is a weaker condition that holds for all positive-mass solutions when the extremality condition~\eqref{eq:extremality} is obeyed.

\section{Near-extremal and near-horizon limits}\label{sec:extremal}

As we observed in Sec.~\ref{sec:general}, in the extremal limit where $r_\pm \rightarrow r_0$, all of our solutions in Eq.~\eqref{eq:solg}, regardless of $q$ value, approach the universal form of the extremal GHS black hole.
However, we may still find interesting differences among these solutions in the NHNE limit.
Recall that in the case of a Reissner-Nordstr\"om black hole in Einstein-Maxwell theory, the near-horizon limit of both the exactly-extremal and the near-extremal black hole is the Bertotti-Robinson geometry of ${\rm AdS}_2 \times S^2$ (in Poincar\'e and Rindler coordinates, respectively) with equal ${\rm AdS}_2$ and $S^2$ scales.
Moreover, Ref.~\cite{Porfyriadis:2021zfb} showed that, in the case of an exactly-extremal GHS black hole, a particular near-horizon limit of the string-frame metric also takes the ${\rm AdS}_2 \times S^2$ form, with the ratio of the length scales of the ${\rm AdS}_2$ and $S^2$ factors depending on $\lambda$.
Here, however, we will find a solution structure for the NHNE limit of Eq.~\eqref{eq:solg} that is much richer.

Let us rewrite $r_\pm = r_0(1\pm \mu \epsilon)$, where $\mu$ and $\epsilon$ are positive parameters, and we take $\epsilon\rightarrow 0$ in the near-extremal limit.
We define new time and radial coordinates $\tau$ and $\rho$,
\be 
\begin{aligned}
t &= \left(\frac{2}{\epsilon}\right)^{1/\gamma} r_0 \tau \\
r &= r_0 \left[1+\frac{\epsilon}{2}\rho^\gamma(1+\mu^2 \rho^{-2\gamma})\right],
\end{aligned}\label{eq:NHNEcoords}
\ee
so that $\epsilon\rightarrow 0$ also corresponds to a simultaneous near-horizon limit, $r\rightarrow r_0$.
Here, we have found it convenient to recast the $\lambda$ parameter via $\gamma = \frac{1+\lambda^2}{1-\lambda^2}$. 
As defined above, and writing $\phi_0 = \bar\phi_0 + \frac{\lambda}{1+\lambda^2} \log(\epsilon/2)$, we find that taking $\epsilon\rightarrow 0$ yields a well defined NHNE limit:
\be 
\begin{aligned}
{\rm d}s^2 \,=&  -r_0^2 \gamma^2 \left(\frac{\rho^\gamma - \mu}{\rho^\gamma+\mu}\right)^{\frac{4\lambda\sqrt{1-q^2+\lambda^2}}{q(1+\lambda^2)}}\frac{(\rho^{2\gamma}-\mu^2)^{2/\gamma}}{\rho^2}{\rm d}\tau^2 
\\ &+ \frac{256 r_0^2 \gamma^2 \mu^4 \rho^{4\gamma-2}}{q^4 (\rho^{2\gamma}-\mu^2)^4 \left[\left(\frac{\rho^\gamma - \mu}{\rho^\gamma+\mu}\right)^{1/q}-\left(\frac{\rho^\gamma - \mu}{\rho^\gamma+\mu}\right)^{-1/q}\right]^4} {\rm d}\rho^2 
\\ &+ \frac{16\mu^2 r_0^2 \rho^{2\gamma}}{q^2 (\rho^{2\gamma} - \mu^2)^2 \left[\left(\frac{\rho^\gamma - \mu}{\rho^\gamma+\mu}\right)^{1/q}-\left(\frac{\rho^\gamma - \mu}{\rho^\gamma+\mu}\right)^{-1/q}\right]^2}{\rm d}\Omega^2 \\
\phi \,=&\; \bar\phi_0 + \frac{\lambda}{1+\lambda^2}\log\left(\frac{\rho^\gamma}{\rho^{2\gamma}-\mu^2}\right) + \frac{\sqrt{1-q^2+\lambda^2}}{q(1+\lambda^2)} \log\left(\frac{\rho^\gamma - \mu}{\rho^\gamma+\mu}\right).
\end{aligned}\label{eq:nearhorizon}
\ee
The geometry in Eq.~\eqref{eq:nearhorizon} represents a two-parameter family of NHNE backgrounds characterized by $q$ and $\mu$ (in addition to the overall scale $r_0$), and on its own represents a new, distinct solution of the EMD equations.

For the particular case of the GHS black hole, $q=1$, Eq.~\eqref{eq:nearhorizon} becomes
\be
\begin{aligned}
{\rm d}s^2 &= r_0^2 \left\{\gamma^2\left[- \frac{(\rho^\gamma-\mu)^2}{\rho^2(\rho^\gamma+\mu)^{2\left(1-\frac{2}{\gamma}\right)}} {\rm d}\tau^2 +\frac{{\rm d}\rho^2}{\rho^2}\right] +{\rm d}\Omega^2\right\} \\
\phi &= \bar\phi_0 + \frac{\lambda}{1+\lambda^2}\log\left[\frac{\rho^\gamma}{(\mu+\rho^\gamma)^2}\right].
\end{aligned}\label{eq:musolution}
\ee
Notably, the NHNE limit factorizes in this case into a product metric of a two-sphere times a warped geometry in the $(\tau,\rho)$ subspace. When $\mu=0$, we obtain the ${\rm AdS}_2 \times S^2$ solution discovered in Ref.~\cite{Porfyriadis:2021zfb},
\be
\begin{aligned}
{\rm d}s^2 &= r_0^2 \left[\gamma^2 \left(-\rho^2 {\rm d}\tau^2 + \frac{{\rm d}\rho^2}{\rho^2}\right)  +  {\rm d}\Omega^2 \right] \\
\phi &= \bar\phi_0 -\frac{\lambda}{1-\lambda^{2}}\log\rho.
\end{aligned}\label{eq:AdS2}
\ee
That is, as discussed in Ref.~\cite{Porfyriadis:2021zfb}, the near-horizon limit of the extremal GHS black hole (in string frame) is an ${\rm AdS}_2 \times S^2$ metric that, unlike the Bertotti-Robinson solution for near-horizon extremal Reissner-Nordstr\"om black holes, has different ${\rm AdS}_2$ and $S^2$ length scales controlled by the dilaton $\lambda$.
What we have found in Eq.~\eqref{eq:musolution} is another remarkable difference from the nondilatonic case. As mentioned above, for standard Einstein-Maxwell theory even near- but sub-extremal Reissner-Nordstr\"om black holes have an ${\rm AdS}_2 \times S^2$ near-horizon limit.
However, for $\lambda\neq 0$, the solution in Eq.~\eqref{eq:musolution} is not diffeomorphic to ${\rm AdS}_2 \times S^2$, as one can confirm by computing the Ricci tensor.

The $q=1$ case of Eq.~\eqref{eq:nearhorizon} was special in that only for this value of $q$ does the angular part of the metric factorize into a distinct $S^2$.
For other values of $q$, we obtain qualitatively different geometries; for example, taking $q=2$ and sending $\rho\rightarrow \left(\frac{\mu}{\rho}\sqrt{\rho^2-r_0^2}\right)^{1/\gamma}$ and $\tau\rightarrow \gamma \tau/(r_0 \mu^\gamma)$, we obtain
\be
{\rm d}s^2 = -\left[\frac{\rho^2}{r_0^2}\left(\frac{\rho^2}{r_0^2}-1\right)\right]^{\frac{\lambda^2-1}{\lambda^2+1}} \left(\frac{\rho-\sqrt{\rho^2-r_0^2}}{\rho+\sqrt{\rho^2-r_0^2}}\right)^{\frac{2\lambda\sqrt{\lambda^2-3}}{1+\lambda^2}} {\rm d}\tau^2 + \frac{\rho^2 r_0^2}{(\rho^2 - r_0^2)^2} {\rm d}\rho^2 + \rho^2 {\rm d}\Omega^2,
\ee
where $\lambda \geq \sqrt{3}$. At large $\rho$, ${\rm d}s^2 \rightarrow -2^{-\frac{4\lambda\sqrt{\lambda^2-3}}{1+\lambda^2}}(\rho/r_0)^n {\rm d}\tau^2 +  (r_0/\rho)^2 {\rm d}\rho^2 +\rho^2 {\rm d}\Omega^2$, where $n=4(\lambda^2-1-\lambda\sqrt{\lambda^2-3})/(1+\lambda^2)$.
That is, the $q=2$ NHNE solution is asymptotically locally a four-dimensional Lifshitz spacetime, reducing to ${\rm AdS}_4$ in the case $\lambda=\sqrt{3}$.

\section{Static horizon perturbations}\label{sec:pert}

In Ref.~\cite{Porfyriadis:2021zfb}, the propagating solutions to the perturbative EMD equations around the ${\rm AdS}_2 \times S^2$ solution in Eq.~\eqref{eq:AdS2} were found.
However, in the static limit, one expects the existence of nonpropagating perturbative solutions as well.
In this section we investigate this question in the s-wave sector.
In particular, we will find the full set of static, spherically symmetric perturbative solutions around Eq.~\eqref{eq:AdS2}.
Ultimately, we will match these solutions to the near-horizon behavior of the full, nonperturbative solutions investigated in Sec.~\ref{sec:general}.

Writing perturbations to the (string-frame) metric $g_{\mu\nu}$, gauge field strength $F_{\mu\nu}$, and dilaton $\phi$ as $h_{\mu\nu}$, $f_{\mu\nu}$, and $\chi$, respectively, the linearized Einstein, Maxwell, and Klein-Gordon equations---$\delta \hat E_\mu^{\;\;\nu} = 0$, $\delta \hat M^\mu = 0$, and $\delta \hat S = 0$---can all be found in Ref.~\cite{Porfyriadis:2021zfb}. 
We restate them here for convenience:
\begin{equation}
\begin{aligned}
\delta \hat E_{\mu}^{\;\;\nu} & =\frac{1}{2}\left(\nabla_{\rho}\nabla_{\mu}h^{\rho\nu}+\nabla^{\rho}\nabla^{\nu}h_{\rho\mu}-\Box h_{\mu}^{\;\;\nu}-\nabla_{\mu}\nabla^{\nu}h_{\rho}^{\;\;\rho}\right)\\
 & \qquad+(2\lambda^{2}-2)\left(\nabla_{\mu}\phi\nabla^{\nu}\chi+\nabla_{\mu}\chi\nabla^{\nu}\phi\right)+2\lambda\nabla_{\mu}\nabla^{\nu}\chi\\
 & \qquad-\lambda\nabla_{\rho}\phi\left(\nabla_{\mu}h^{\nu\rho}+\nabla^{\nu}h_{\mu}^{\;\;\rho}-\nabla^{\rho}h_{\mu}^{\;\;\nu}\right)+\lambda h_{\mu}^{\;\;\nu}\left[\Box\phi-2\lambda(\nabla\phi)^{2}\right]\\
 & \qquad+\lambda\delta_{\mu}^{\nu}\left[-h^{\rho\sigma}\nabla_{\rho}\nabla_{\sigma}\phi-\nabla^{\sigma}\phi\left(\nabla^{\rho}h_{\rho\sigma}-\frac{1}{2}\nabla_{\sigma}h_{\rho}^{\;\;\rho}\right)+2\lambda h^{\rho\sigma}\nabla_{\rho}\phi\nabla_{\sigma}\phi\right]\\
  & \qquad+\lambda\delta_{\mu}^{\nu}\left(\Box\chi-4\lambda\nabla^{\rho}\phi\nabla_{\rho}\chi\right)\\
 & \qquad-\frac{1}{2}\delta_{\mu}^{\nu}F_{\rho\alpha}F_{\sigma}^{\;\;\alpha}h^{\rho\sigma}+F_{\mu\rho}F^{\nu\sigma}h_{\;\;\sigma}^{\rho}+\frac{1}{4}F_{\rho\sigma}F^{\rho\sigma}h_{\mu}^{\;\;\nu}-F_{\mu\rho}f^{\nu\rho}-F^{\nu\rho}f_{\mu\rho}+\frac{1}{2}\delta_{\mu}^{\nu}F^{\rho\sigma}f_{\rho\sigma},\\
\delta \hat M^{\mu} & =\nabla_{\nu}f^{\nu\mu}-2\lambda f^{\nu\mu}\nabla_{\nu}\phi-h^{\nu\rho}\nabla_{\rho}F_{\nu}^{\;\;\mu}-F_{\rho}^{\;\;\mu}\nabla_{\nu}h^{\nu\rho}-F^{\nu\rho}\nabla_{\nu}h_{\rho}^{\;\;\mu}\\
 & \qquad+\frac{1}{2}F^{\nu\mu}\nabla_{\nu}h_{\rho}^{\;\;\rho}+2\lambda F^{\nu\mu}h_{\nu\rho}\nabla^{\rho}\phi-2\lambda F^{\rho\mu}\nabla_{\rho}\chi, 
  \end{aligned}
 \end{equation}
 
 \begin{equation*}
\hspace{-24.5mm} \begin{aligned}
\delta \hat S & =(3\lambda^{2}-1)\left(\nabla_{\mu}\phi\nabla^{\mu}h_{\nu}^{\;\;\nu}-2\nabla_{\mu}\phi\nabla_{\nu}h^{\mu\nu}-2h^{\mu\nu}\nabla_{\mu}\nabla_{\nu}\phi+2\lambda h^{\mu\nu}\nabla_{\mu}\phi\nabla_{\nu}\phi\right)\\
&\qquad +2(3\lambda^{2}-1)\left(\Box\chi-2\lambda\nabla_{\mu}\phi\nabla^{\mu}\chi\right)\\
 & \qquad+\lambda\nabla_{\mu}\nabla_{\nu}h^{\mu\nu}-\lambda\Box h_{\mu}^{\;\;\mu}-\lambda h^{\mu\nu}R_{\mu\nu}-\lambda F^{\mu\nu}f_{\mu\nu}+\lambda F_{\mu\rho}F_{\nu}^{\;\;\rho}h^{\mu\nu}.
\end{aligned}\label{eq:perteqs}
\end{equation*}
For the perturbations, we take the ansatz,
\be
\begin{aligned}
h_{\mu\nu}{\rm d}x^\mu {\rm d}x^\nu &= \gamma^2 r_0^2\left[\rho^2 H_0(\rho) {\rm d}\tau^2 + 2 H_1(\rho){\rm d}\tau {\rm d}\rho + \rho^{-2} H_2(\rho){\rm d}\rho^2\right] +  r_0^2 K(\rho){\rm d}\Omega^2 \\ 
f &= \sqrt{\frac{2}{1+\lambda^2}}r_0 v(\rho) \sin \theta \,{\rm d}\theta\wedge {\rm d}\varphi \\ 
\chi &= \chi(\rho).
\end{aligned}\label{eq:pertansatz}
\ee
Acting with a diffeomorphism $x^\mu \rightarrow x^\mu + \xi^\mu$, where
\be
\xi = \left(\int^\rho \frac{2H_1(\hat\rho)}{\hat\rho^2}{\rm d}\hat\rho\right)\partial_\tau - \rho\left(\int^\rho \frac{H_2(\hat\rho)}{2\hat\rho} {\rm d}\hat\rho\right) \partial_\rho,
\ee
eliminates $H_1(\rho)$ and $H_2(\rho)$ from the ansatz~\eqref{eq:pertansatz}.\footnote{A general diffeomorphism $\xi$ generates a perturbation to the metric $\delta h_{\mu\nu} = {\cal L}_\xi g_{\mu\nu} = \nabla_\mu \xi_\nu + \nabla_\nu \xi_\mu$, field strength $\delta f_{\mu\nu}={\cal L}_\xi F_{\mu\nu}=\xi^\sigma \nabla_\sigma F_{\mu\nu}+F_{\mu\sigma} \nabla_\nu\xi^\sigma-F_{\nu\sigma}\nabla_\mu\xi^\sigma$, and dilaton $\delta\chi = {\cal L}_\xi \phi = \xi^\mu \nabla_\mu \phi$.}
We can solve the Klein-Gordon equation $\delta\hat S = 0$ algebraically for $v(\rho)$,
\be
\begin{aligned}
v(\rho) &= \frac{1-\lambda^2}{2}\left[K(\rho) - \rho K'(\rho) - \frac{1-\lambda^2}{1+\lambda^2}\rho^2 K''(\rho)\right]  \\ & \;\;\;\; + \frac{1-\lambda^2}{2(1+\lambda^2)}\rho \left[H_0'(\rho) +\frac{1-\lambda^2}{2}\rho H_0''(\rho)\right] \\& \;\;\;\; -\frac{(1-\lambda^2)(1-3\lambda^2)}{2\lambda(1+\lambda^2)} \rho \left[2 \chi'(\rho) + (1-\lambda^2)\rho \chi''(\rho) \right].
\end{aligned}
\ee
We then find that $\delta \hat E_\tau^{\;\;\tau} - \delta \hat E_\rho^{\;\;\rho} - 2\delta\hat E_\theta^{\;\;\theta}=0$ yields a first-order differential equation for $\chi'(\rho)$,
\be
\chi'(\rho) = c_0 \rho^{-\frac{2}{1-\lambda^2}} + \frac{\lambda}{2(1-\lambda^2)}H_0'(\rho),
\ee
where $c_0$ is an integration constant.
Subsequently, we find that $\delta \hat E_\tau^{\;\;\tau} - \frac{1+2\lambda^2}{1+4\lambda^2} \delta E_\rho^{\;\;\rho}=0$ algebraically yields $H_0'(\rho)$,
\be
H_0'(\rho) = \frac{K(\rho)}{\rho} -\frac{1-\lambda^2}{2\lambda^2(1+\lambda^2)^2}\left[8c_0 \lambda^3(1+\lambda^2)\rho^{-\frac{2}{1-\lambda^2}} - 2\lambda^2(\lambda^2+2)K'(\rho)+(1-\lambda^2)\rho K''(\rho)\right].
\ee
Finally, defining $\bar K(\rho) = \rho^{\frac{2-\lambda^2}{1-\lambda^2}}K'(\rho)$, we find that the remaining equation of motion $\delta \hat E_\tau^{\;\;\tau}=0$ becomes
\be
\bar K''(\rho) - \frac{(\lambda^2+2)(2\lambda^2+1)}{(1-\lambda^2)^2\rho^2}\bar K(\rho) = 0 .
\ee
The general solution is
\be
\bar K(\rho) = c_+ \rho^{\frac{\lambda^2+2}{1-\lambda^2}} + c_- \rho^{-\frac{2\lambda^2+1}{1-\lambda^2}}.
\ee
Defining $e_0 = -\lambda c_0/\gamma^2$, $e_+ = c_+/\gamma$, and $e_- = -c_-/2\gamma$, along with new integration constants $B$, $\chi_0$, and $\Phi_0$, we compute the integrals to find $H_0$, $K$, and $\chi$, finding the general solution,
\be
\begin{aligned}
K(\rho)&= e_+ \rho^\gamma + e_- \rho^{-2\gamma} + 2\Phi_0 \\
H_0(\rho) &= -4e_0 \rho^{-\gamma} + \frac{2}{\gamma}e_+ \rho^\gamma - \frac{1+2\gamma}{\gamma(1-\gamma)} e_- \rho^{-2\gamma} + 2 \Phi_0 \log\rho + B \\
\chi(\rho) &= \chi_0 + \frac{1}{\lambda}\left[e_0 \rho^{-\gamma} -\frac{1-\gamma}{2\gamma}e_+ \rho^\gamma + \frac{1+2\gamma}{4\gamma} e_- \rho^{-2\gamma}  - \frac{1-\gamma}{2}\Phi_0 \log \rho \right] \\
v(\rho) &= \Phi_0.
\end{aligned} \label{eq:pertsol}
\ee
The constant $\chi_0$ simply represents a perturbation of the overall offset $\phi_0$ of the dilaton; since the dilaton couples purely derivatively in the string-frame equations of motion given in Eqs.~\eqref{eq:ein}, \eqref{eq:max}, and \eqref{eq:KG}, this offset is not physical.
Similarly, the constant $B$ appearing in $H_0$ is pure gauge and can be removed via a diffeomorphism $\xi = B\rho \partial_\rho/2$.
We therefore drop $B$ and $\chi_0$.
The constants $e_0$, $e_+$, $e_-$, and $\Phi_0$, in contrast, are physically meaningful.

The meaning of $\Phi_0$ is easy to understand as a rescaling of Eq.~\eqref{eq:AdS2} by $r_0\to r_0+\delta r_0$. Indeed, the perturbative solution
\be
\begin{aligned}
	h_{\mu\nu}{\rm d}x^\mu {\rm d}x^\nu &= 2r_0\delta r_0 \left[\gamma^2 \left(-\rho^2 {\rm d}\tau^2 + \frac{{\rm d}\rho^2}{\rho^2}\right)  +  {\rm d}\Omega^2 \right] \\
	f &= \sqrt{{2\over 1+\lambda^2}} \, \delta r_0 \sin\theta\,{\rm d}\theta\wedge {\rm d}\varphi \\
	\chi &= 0,
\end{aligned}
\ee
may be brought to the gauge used in this section,
\be 
\begin{aligned}
	h_{\mu\nu}{\rm d}x^\mu {\rm d}x^\nu &= 2\gamma^2 r_0 \delta r_0 \rho^2 \log\rho {\rm d}\tau^2 +2 r_0\delta r_0{\rm d}\Omega^2 \\
	f &= \sqrt{\frac{2}{1+\lambda^2}}\delta r_0 \sin\theta\,{\rm d}\theta\wedge {\rm d}\varphi \\
	\chi &= \frac{\lambda}{1-\lambda^2}(\delta r_0 /r_0) \log\rho ,
\end{aligned}\label{eq:canonicalpert}
\ee
by acting with the diffeomorphism generated by $\xi = -(\delta r_0 /r_0)\left(\tau \partial_\tau+\rho \log \rho \partial_\rho \right)$, so that comparing with Eqs.~\eqref{eq:pertansatz} and \eqref{eq:pertsol} we see that the rescaling of Eq.~\eqref{eq:AdS2} corresponds to the $\Phi_0=\delta r_0/r_0$ solution. 
This leaves us with the more interesting solutions $e_+$, $e_0$, $e_-$, which are discussed in the next section.

\section{Anabasis}\label{sec:anabasis}

In this section we show that one may interpret the physical perturbative near-horizon solutions parameterized by $e_+$, $e_0$, $e_-$ in Eq.~\eqref{eq:pertsol} as encoding the first step toward building various nonlinear solutions outward from the vicinity of $r=r_+$.
That is, we turn to the problem of anabasis for the ${\rm AdS}_2 \times S^2$ geometry~\eqref{eq:AdS2} in EMD gravity, in analogy with the pure Einstein-Maxwell anabasis considered in Ref.~\cite{Hadar:2020kry}.
As we will see, the mode $e_+$ is the one responsible for breaking away from the near-horizon  ${\rm AdS}_2$ region, while the modes $e_0$ and $e_-$ parameterize possible deviations from extremality of the resulting asymptotically flat solutions.

Since the ${\rm AdS}_2 \times S^2$ geometry~\eqref{eq:AdS2} is the near-horizon limit of the extremal GHS black hole~\eqref{eq:extremal}, we may read off an anabasis perturbation of ${\rm AdS}_2 \times S^2$ toward extreme GHS from the first correction to this limit. The limit, first identified in Ref.~\cite{Porfyriadis:2021zfb}, is given by Eq.~\eqref{eq:NHNEcoords} with $\mu=0$. At leading order, this limit produces the ${\rm AdS}_2 \times S^2$ geometry~\eqref{eq:AdS2}, and the first correction is a perturbative solution around it:
\be
\begin{aligned}
	h_{\mu\nu}{\rm d}x^\mu {\rm d}x^\nu &= r_0^2 \left(\gamma\rho^{\gamma+2}{\rm d}\tau^2 + {\gamma^2}\rho^{\gamma-2}{\rm d}\rho^2 + \rho^\gamma {\rm d}\Omega^2\right) \\
	\chi &= \frac{\lambda}{2(1+\lambda^2)}\rho^\gamma.
\end{aligned}
\ee
Adjusting the gauge by acting with an infinitesimal diffeomorphism along $\xi = -\frac{1}{2\gamma} \rho^{\gamma+1} \partial_\rho$ brings the above perturbation into the form matching our gauge of Sec.~\ref{sec:pert}:
\be
\begin{aligned}
	h_{\mu\nu}{\rm d}x^\mu {\rm d}x^\nu &= r_0^2 \left(2\gamma\rho^{\gamma+2}{\rm d}\tau^2 + \rho^\gamma {\rm d}\Omega^2\right) \\
	\chi &= \frac{\lambda}{1+\lambda^2}\rho^\gamma.
\end{aligned}
\ee
Comparing with Eqs.~\eqref{eq:pertansatz} and \eqref{eq:pertsol}, we see that this is the $e_+$ mode, with $e_+$ set to unity. That is, we see that the $e_+$ mode begins to implement the anabasis from the ${\rm AdS}_2 \times S^2$ near-horizon throat toward the asymptotically flat extreme GHS black hole.

Turning on $e_0$ or $e_-$, in addition to $e_+$, leads to anabasis to near-extreme spacetimes in the general family described by Eq.~\eqref{eq:solg}. This situation is similar to that in pure Einstein-Maxwell theory, where anabasis from Bertotti-Robinson may lead to both extreme and near-extreme Reissner-Nordstr\"om. However, unlike the nondilatonic case, here ${\rm AdS}_2 \times S^2$ arises in the near-horizon limit only when the near-extreme limit is taken at appropriate faster rates than the one considered in Sec.~\ref{sec:extremal}, where as we saw taking the two limits at the same rate produces the geometry in Eq.~\eqref{eq:nearhorizon}, rather than ${\rm AdS}_2 \times S^2$. 

The $e_0$ mode may be identified by taking the near-horizon limit in Eq.~\eqref{eq:NHNEcoords} together with a near-extreme limit given by $r_\pm=r_0(1\pm\bar\mu\epsilon^2)$, keeping $\bar{\mu}$ fixed (instead of the limit $r_\pm=r_0(1\pm\mu\epsilon)$ for finite $\mu$ considered in Sec.~\ref{sec:extremal}). In other words, this is a limit of the general solution \eqref{eq:solg} such that $r-r_0 \sim O(\epsilon)$ and $r_+ - r_- \sim O(\epsilon^2)$. Indeed, taking the $\epsilon\to 0$ limit produces the ${\rm AdS}_2 \times S^2$ solution \eqref{eq:AdS2} at leading order. The first-order correction around this limit is given by
\be
\begin{aligned}
	h_{\mu\nu}{\rm d}x^\mu {\rm d}x^\nu &= r_0^2 \left[ \left(\gamma\rho^{\gamma}+\frac{8\lambda\gamma\sqrt{1-q^2+\lambda^2}}{q(1-\lambda^2)}\bar\mu \rho^{-\gamma}\right)\rho^2{\rm d}\tau^2 +  {\gamma^2}\rho^{\gamma-2}{\rm d}\rho^2 + \rho^\gamma{\rm d}\Omega^2 \right] \\
	\chi &= \frac{\lambda}{2(1+\lambda^2)}\rho^\gamma - \frac{2\sqrt{1-q^2+\lambda^2}}{q(1+\lambda^2)} \bar\mu \rho^{-\gamma},
\end{aligned}
\ee
which after adjusting the gauge along $\xi = -\frac{1}{2\gamma} \rho^{\gamma+1} \partial_\rho$ becomes
\be
\begin{aligned}
	h_{\mu\nu}{\rm d}x^\mu {\rm d}x^\nu &= r_0^2 \left[ \left(2\gamma\rho^{\gamma}+\frac{8\lambda\gamma\sqrt{1-q^2+\lambda^2}}{q(1-\lambda^2)}\bar\mu \rho^{-\gamma}\right)\rho^2{\rm d}\tau^2 + \rho^\gamma{\rm d}\Omega^2 \right] \\
	\chi &= \frac{\lambda}{1+\lambda^2}\rho^\gamma - \frac{2\sqrt{1-q^2+\lambda^2}}{q(1+\lambda^2)} \bar\mu \rho^{-\gamma}.
\end{aligned}
\ee
Comparing with Eqs.~\eqref{eq:pertansatz} and \eqref{eq:pertsol}, we see that this is the solution normalized by $e_+=1$ with
\be\label{eq:e0 anabasis}
e_0=-\frac{2 \bar\mu \lambda \sqrt{1-q^2 + \lambda^2}}{q(1+\lambda^2)}.
\ee
That is, we see that in anabasis from ${\rm AdS}_2 \times S^2$, the $e_0$ mode  may be used to build a near-extreme member of the family \eqref{eq:solg} with deviation from extremality given by $(r_+ - r_-)/r_0 =2\bar{\mu}\epsilon^2=-e_0q \epsilon^2 (1+\lambda^2)/(\lambda \sqrt{1-q^2 + \lambda^2})$. Notice that this deviation from extremality is controlled by $e_0$ but is also parameterized by an \emph{arbitrary} choice of $q$, so long as $q^2\neq 1+\lambda^2$. This is of course linked to the attractor role that the extreme GHS solution \eqref{eq:extremal} plays in the space of all $q$-deformed solutions \eqref{eq:solg} near extremality.

When $q^2= 1+\lambda^2$, there exists an alternative, somewhat slower, limit toward extremality that also furnishes an ${\rm AdS}_2 \times S^2$ geometry in the near-horizon region. This is the limit of the general solution \eqref{eq:solg} with $r-r_0\sim O(\epsilon)$ and $r_+ - r_-\sim O(\epsilon^{3/2})$. Specifically, taking the near-horizon limit in Eq.~\eqref{eq:NHNEcoords} together with a near-extreme limit given by $r_\pm=r_0(1\pm\bar{\bar{\mu}}\epsilon^{3/2})$, keeping $\bar{\bar{\mu}}$ fixed, yields the ${\rm AdS}_2 \times S^2$ solution \eqref{eq:AdS2} at leading order, with the perturbative $O(\epsilon)$ solution around it  given by
\be
\begin{aligned}
	h_{\mu\nu}{\rm d}x^\mu {\rm d}x^\nu &= r_0^2 \Bigg[ \left(\gamma\rho^{\gamma}+2\gamma\bar{\bar{\mu}}^2 \rho^{-2\gamma}\right)\rho^2{\rm d}\tau^2 \\ &
	\qquad\;\;\; +  
	\left(\gamma^2\rho^{\gamma-2}+\frac{4 \gamma(\gamma-1)}{3}\bar{\bar{\mu}}^2\rho^{-2\gamma-2}\right){\rm d}\rho^2 + \left(\rho^\gamma+\frac{2(\gamma-1)}{3\gamma}\bar{\bar{\mu}}^2\rho^{-2\gamma}\right){\rm d}\Omega^2 \Bigg] \\
	\chi &= \frac{\lambda}{2(1+\lambda^2)}\rho^\gamma+
	\frac{{\Bar{\Bar{\mu}}}^2\lambda}{1+\lambda^2}\rho^{-2\gamma}.
\end{aligned}
\ee
Adjusting the gauge along $\xi = -\frac{1}{\gamma}\rho \left(\frac{1}{2}\rho^\gamma + \frac{\gamma-1}{3\gamma}\bar{\bar{\mu}}^2 \rho^{-2\gamma}\right) \partial_\rho$, this solution becomes
\be
\begin{aligned}
	h_{\mu\nu}{\rm d}x^\mu {\rm d}x^\nu &= r_0^2 \left[ \left(2\gamma\rho^{\gamma}+\frac{2(1+2\gamma)}{3}\Bar{\Bar{\mu}}^2 \rho^{-2\gamma}\right)\rho^2{\rm d}\tau^2 + \left(\rho^\gamma+\frac{2(\gamma-1)}{3\gamma}\bar{\bar{\mu}}^2\rho^{-2\gamma}\right){\rm d}\Omega^2 \right] \\
	\chi &= \frac{\lambda}{1+\lambda^2}\rho^\gamma +\frac{(1+\gamma)(1+2\gamma)}{6\gamma^2}\lambda\bar{\bar{\mu}}^2\rho^{-2\gamma},
\end{aligned}
\ee
so that comparing with Eqs.~\eqref{eq:pertansatz} and \eqref{eq:pertsol}, we see that this is the solution normalized by $e_+=1$ with
\be
e_- = \frac{2(\gamma-1)}{3\gamma}\Bar{\Bar{\mu}}^2.
\ee
That is to say, we see that in anabasis from ${\rm AdS}_2 \times S^2$, the $e_-$ mode  may be used to build a near-extreme member of the family \eqref{eq:solg} with $q^2=1+\lambda^2$ and deviation from extremality given by $(r_+ - r_-)/r_0 =2\bar{\bar{\mu}}\epsilon^{3/2}=\sqrt{6\gamma e_-/(\gamma-1)}\epsilon^{3/2}$.

\section{Conclusions}\label{sec:conclusions}

In this paper we have presented the general three-parameter family of static, spherically symmetric, magnetic solutions to the EMD equations in string frame. We have paid particular attention to the extremal limit and its near-horizon geometries. Remarkably, the exactly-extremal spacetime coincides with the one-parameter extreme GHS black hole, which acts as an attractor and whose near-horizon geometry in string frame is ${\rm AdS}_2 \times S^2$~\cite{Porfyriadis:2021zfb}. On the other hand, for a near-extremal GHS black hole, we have identified a new warped ${\rm AdS}_2$ near-horizon geometry~\eqref{eq:musolution}. Moreover, a full two-parameter family of near-horizon solutions to our $q$-parameterized family of more general objects is given in Eq.~\eqref{eq:nearhorizon}. 

We have also studied in detail the problem of linear deformations of ${\rm AdS}_2 \times S^2$ in EMD gravity and the associated phenomenon of anabasis that leads to nonlinear solutions with different asymptotics. Specifically, we have found that static linear deformations of ${\rm AdS}_2 \times S^2$ are characterized by four parameters, of which three are related to anabasis. Specifically, one of the perturbations leads to the extreme GHS black hole, while the other two add deviation from extremality. It is interesting that in EMD, unlike the case of pure Einstein-Maxwell theory~\cite{Hadar:2020kry}, we found that the linear anabasis perturbation in Eq.~\eqref{eq:e0 anabasis} does not completely fix the deviation from extremality, which is further parameterized by an additional arbitrary parameter that breaks the degeneracy of the extreme GHS attractor solution in EMD.

This paper leaves compelling avenues for future work.
While the general solution, in Einstein frame, describes pointlike objects when $q\neq 1$, these solutions would nonetheless be useful in a string theory context to describe an object of general $(M,P,D)$ charges outside the singular region (just as Reissner-Nordstr\"om describes the metric sufficiently far outside of a point charge in Einstein-Maxwell theory).
It would be extremely interesting to understand these new charged solutions, and their deformations under higher-derivative corrections, in the context of the weak gravity conjecture~\cite{Arkani-Hamed:2021ajd}.
Moreover, for $q>2$, we have seen that the string-frame metric is regular at $r=r_+$, but that the object has divergent area; a full exploration of the physical properties of these solutions remains to be pursued.
Finally, the GHS black hole has thermodynamic properties that differ markedly from the Reissner-Nordstr\"om case~\cite{GHS}.
A thermodynamic analysis of our $q$-parameterized solutions, as well as an investigation of the physics behind the attractor behavior of the extreme GHS solution, certainly warrants further study.

\medskip

\begin{center}
{\bf Acknowledgments}
\end{center}
\noindent
We thank Gary Horowitz and Andy Strominger for useful discussions and comments.
G.N.R. is supported at the Kavli Institute for Theoretical Physics by the Simons Foundation (Grant No. 216179) and the National Science Foundation (Grant No. NSF PHY-1748958) and at the University of California, Santa Barbara by the
Fundamental Physics Fellowship.

\bibliographystyle{utphys-modified}
\bibliography{EMDsolutions}

\end{document}